\documentclass[letterpaper]{article}
\usepackage{aaai}
\usepackage{times}
\usepackage{helvet}
\usepackage{courier}
\usepackage{graphicx}
\usepackage{booktabs}
\usepackage{float}
\usepackage{longtable}
\usepackage{xurl}

\begin{document}
%
\title{TWikiL - the Twitter Wikipedia Link Dataset}

\author{Florian Meier\\
Department of Communication and Psychology\\
Aalborg University, Copenhagen\\
Copenhagen, Denmark\\
fmeier@ikp.aau.dk\\
}



\maketitle
\begin{abstract}
Recent research has shown how strongly connected Wikipedia and other web applications are. For example, search engines rely heavily on surfacing Wikipedia links to satisfy their users' information needs and volunteer-created Wikipedia content frequently gets re-used on social media platforms like Reddit. However, publicly accessible datasets that enable researchers to study the interrelationship between Wikipedia and other platforms are sparse. In addition to that, most studies only focus on certain points in time and do not consider the historical perspective. To begin solving these problems, we developed \texttt{TWikiL}, the Twitter Wikipedia Link Dataset, which contains all Wikipedia links posted on Twitter in the period 2006 to January 2021. We extract Wikipedia links from Tweets and enrich the referenced articles with their respective Wikidata identifiers and Wikipedia topic categories. This makes the dataset immediately useful for a large range of scholarly use cases. In this paper, we describe the data collection process, perform an initial exploratory analysis and present a comprehensive overview of how this dataset can be useful for the research community.
\end{abstract}

\section{Introduction}
\label{sec:introduction}
Twitter and Wikipedia are among the most visited and researched websites in the world. While the platforms have been mostly studied in isolation, recent research breaks this pattern and we begin to realize that especially Wikipedia is more than just the largest knowledge repository on the web, but has strong relationships and interdependencies with other web applications. Researchers argue that it lies at the center of a social media or information ecology or to quote Vincent et al.: "Wikipedia matters outside of Wikipedia" \cite{Vincent2019}. And not only do we see Wikipedia having impact on the digital world, there is also evidence for Wikipedia having direct effect on e.g. real-world economic outcomes by influencing which cities tourists pick for an over night stay \cite{Hinnosaar2019} or even what type of research is being done \cite{Thompson2018}. 

First and foremost, however, Wikipedia is significantly shaping our digital experience. Most state-of-the-art machine learning algorithms are trained on Wikipedia content \cite{Devlin2019,Cambazoglu2021}, millions of searchers are dependent on Google surfacing Wikipedia links to satisfy their information needs \cite{McMahon2017,Vincent2019,Vincent2021}, tech giants are using human workforce and Wikipedia for fact checking on their platforms \cite{Flynn2017,Perez2020}, Wikipedia links are among the most prominent URLs on community question answering sites like Reddit or Stack Overflow \cite{Gomez2013,Vincent2018}, voice-operated virtual assistants like \textit{Siri} and \textit{Alexa} rely heavily on Wikipedia \cite{Vincent2021} and for many users Wikipedia is a stepping stone or gateway to the larger web itself \cite{Piccardi2021}. 
While Wikipedia is providing a lot of value to other platforms, less value is flowing back to Wikipedia in return. Despite all this exposure, Wikipedia is not seeing an increase in page visits, a growing editor base and edits, or monetary donations. A fact that McMahon termed \textit{the paradox of reuse} \cite{McMahon2017}. Recently, Wikimedia took the initiative to address this paradox and bring more balance in this value flow by launching the Wikimedia Enterprise APIs\footnote{https://enterprise.wikimedia.com/}, a set of API services focused on high-volume for-profit use cases like the ones mentioned above \cite{WikimediaEnterprise2021}.

Ultimately, we are only beginning to see how strongly certain social media platforms depend on each other, and what dynamics between web applications are at play. To shed more light on these dynamics and allow researchers to study what value Twitter has for Wikipedia and vice-versa, we developed \texttt{TWikiL}, a dataset containing all Tweets with Wikipedia links posted between 2006 and January 2021, that captures the relationship and mutual connection of the largest online encyclopedia and one of the most prominent social media platforms. \texttt{TWikiL} data collection was made possible via a Twitter Academic Research access grant and will be released in two versions: \texttt{TWikiL raw}, which is a list of Tweet IDs with links to all URLs in the domain \textit{wikipedia.org}, and \texttt{TWikiL curated}, a database covering all Wikipedia \textit{article} links, enriched with Wikidata IDs for the concepts referenced in these articles and the Wikipedia topic categories they belong to. Our initial exploratory analysis shows an increasing trend in the number of article links posted on Twitter over time, with links to English and Japanese language editions dominating. Moreover, users mostly share links related to \textit{Culture} concepts with links to biographies representing almost one third of all links. Our initial analysis also reveals many different propagation patterns for concepts and shows that bots are likely to play a significant role in the diffusion of Wikipedia links.

We believe that \texttt{TWikiL} offers a broad range of scholarly use cases and could help answer research questions like: \textit{Do biases, with respect to gender or race prominent in Wikipedia article coverage also diffuse to other platforms?} or \textit{How are Wikipedia links used in Twitter conversations and discussions?}. Finally, due to its unique nature covering the evolution of the relationship since Twitter's very first years, this dataset can also be considered a contribution to studies on the history of the web \cite{Bruegger2018}. Before we present the data collection process in detail, we present relevant related work that inspired \texttt{TWikiL}'s creation.

\section{Related Work}
\label{sec:related_work}
Our work was mostly inspired by previous studies that investigate Wikipedia's role outside of Wikipedia. This includes studies on (1) Wikipedia article text re-use, especially for for-profit content farming purposes (2) Wikipedia's connection to large-scale information technologies like search engines and (3) Wikipedia's role in the broader social media or information ecology of the web. We review these areas accordingly.

\subsection{Wikipedia Text Reuse}
Although there are no direct rules that per se speak against duplicating Wikipedia content, limitations for commercial re-use of Wikipedia content exist \cite{WikipediaContent2021}. However, one can see that Wikipedia gets duplicated frequently and falls victim to content farming, where Wikipedia text gets copied to web pages with advertisements with the aim of monetizing it's user generated content. In an early study on replicated web content, Ardi and Heidemann found 40 sites (136 K pages) that copy text from Wikipedia with 86\% of these sites doing this for profit \cite{Ardi2014}. In a similar vein, Alshomary and colleagues studied Wikipedia text reuse within (inside Wikipedia) and without (on the web) \cite{Alshomary2019}. In total, they found 70 times more instances of text reuse within Wikipedia then outside the platform. Outside Wikipedia they identify 4,898 websites for a total 1.6 million reuse cases. Their conservative, lower-bound, estimate of ad-revenue extrapolated to the whole web finds that this text-reuse could be responsible for 5.5 M USD monthly advertising revenue, which is around 72\% of Wikipedia's yearly fundraising returns \cite{Alshomary2019}.

\subsection{Wikipedia and Search Engines}
Among the first to study the interdependence between Wikipedia and search engines were McMahon et al. who looked into the relationship between Wikipedia and Google \cite{McMahon2017}. They performed two controlled search behavior experiments in which a browser extension implemented three experimental conditions to investigate how removing Wikipedia links and/or the Wikipedia Knowledge Graph asset would effect the user's search experience. Their results show that Google becomes a worse search engine for many queries and creates a negative search user experience if Wikipedia content gets removed from search engine result pages (SERPs). While they argue that Wikipedia and Google form a mutually beneficial symbiosis, as surfacing Wikipedia links on Google also leads to more traffic to Wikipedia, newer technologies like the knowledge graph which directly satisfy information needs without the user having to actively visit Wikipedia jeopardize this relationship. McMahon et al. call this the \textit{paradox of reuse} which describes the situation that while more and more Wikipedia content gets surfaced by search engines like Google, less people actually visit Wikipedia itself, which leads to less growth of the editor community and a decline in user generated content through edits \cite{McMahon2017}. Vincent and colleagues extend McMahon et al.'s work by performing a rigorous audit of Google's algorithm and how strongly it relies on user generated content, including content from Wikipedia \cite{Vincent2019}. They observe that for some query types, Wikipedia links appear in over 80\% of SERPs which leads them to the conclusion that Wikipedia's user generated content is invaluable to Google and that no other website in the world is as dependent on Wikipedia as Google is \cite{Vincent2019}. Most recently, Vincent and Hecht go yet one step further and include mobile devices and the two additional search engines, Bing and DuckDuckGo, in a deeper investigation of the importance of Wikipedia's content for SERPs \cite{Vincent2021}. Via a search algorithm audit study, they again prove that Wikipedia content has much influence outside of Wikipedia and that all search engines --- not only Google --- rely to a large degree on the encyclopedia's content, highlighting again how important user generated content by volunteers can be. 

\subsection{Wikipedia and Social Media Ecology}
Finally, research showed that Wikipedia's influence is not limited to search engines, but has impact on the broader ecology of social media platforms that it is situated in \cite{Gomez2013,Moyer2015,Vincent2018}. Gómez, Cleary and Singer studied link sharing on Stack Overflow finding that Wikipedia links account for around 5\% of URLs in their sample \cite{Gomez2013}. They argue that there is a class of posts (foundational topics in computer science) that almost exclusively relies on Wikipedia links. Moyer et al. determine the influence that linking Wikipedia articles in Reddit posts has on Wikipedia pageview statistics \cite{Moyer2015}. On the subreddit \textit{/r/todayilearned} (TIL), Wikipedia article URLs contribute the largest share of links in posts. Through PCA of timeseries data they find that Wikipedia article links posted on Reddit clearly contribute to higher pageviews of these articles on Wikipedia. 
Vincent, Johnson and Hecht, directly influenced by the works of Gómez et al. and Moyer et al., study what role volunteer-created content from Wikipedia has for Stack Overflow and Reddit \cite{Vincent2018}.
On the one hand, they discover that Wikipedia does not only increase these platforms user's experience, but also creates monetary value as Wikipedia is responsible for \$1.7 million in ad revenues for the two platforms. On the other hand, they find strong evidence for the \textit{paradox of reuse} given the fact that Wikipedia-linking posts were receiving a lot of attention on Reddit and Stack Overflow, but this did not result in an increase in edits or pageviews on Wikipedia. As a solution, they suggest design interventions to increase the mutual value of Wikipedia and Stack Overflow/Reddit \cite{Vincent2018}.

To sum up, there is an increasing body of work that situates Wikipedia in the center of the web's information ecology with multiple platforms being directly dependent on Wikipedia and vice versa. To add to that body of research and also being able to characterise the historical evolution of the dependency between Twitter and Wikipedia, we developed \texttt{TWikiL}.

\section{Dataset Development and Description}
\label{sec:dataset_development}

\texttt{TWikiL} includes all Wikipedia links posted on Twitter from the very first link posted in July 2006 to January 2021. 
The dataset does not include links to non-Wikipedia projects such as Wikimedia Commons, Wiktionary or Wikiquote, but prioritizes links to Wikipedia pages. The dataset comes in two versions: (1) \texttt{TWikiL raw} which is a CSV file containing the IDs of all Tweets that link to Wikipedia, i.e. that are in the domain \textit{wikipedia.org} and (2) \texttt{TWikiL curated}, a SQLite database which contains a curated version of \texttt{TWikiL} for links to Wikipedia article pages or other types of pages with a Wikidata entry that are not main pages. This means the curated version does not contain links to talk\footnote{https://en.wikipedia.org/wiki/Talk:Twitter}, history\footnote{https://en.wikipedia.org/w/index.php?title=Talk:Twitter\&action=history}, user\footnote{https://en.wikipedia.org/wiki/User:[user\_name]}, main\footnote{https://en.wikipedia.org/wiki/Main\_Page} or similar pages which are otherwise part of the domain \textit{wikipedia.org} but do not directly link to articles or do not have a Wikidata identifier. However, these are still part of \texttt{TWikiL raw}. \texttt{TWikiL curated} was mostly created for convenience sake to make it easier to work with the dataset, especially when deciding which of Tweets should be hydrated, i.e. retrieved as Tweet objects with full text etc. via the Twitter API, for further analysis. Both versions of the dataset are accessible and can be downloaded via Zenodo \cite{Meier2022}.

In what follows, we describe the data collection process, how we performed additional processing steps, which limitations the dataset faces and finally express our considerations regarding ethics and the FAIR principles for scientific data management.

\subsection{Data Collection and Processing}
\texttt{TWikiL} data collection was made possible via Twitter's Academic Research access level which grants access to full-archive search via the recently introduced v2 API. This API access allows for retrieving up to 10 million Tweets per month, the use of advanced search operators and the retrieval of more detailed data objects for Tweets including entity objects that provide additional information about URLs like their expanded version \cite{TwitterDataDictionary2021}. We used the fairly simple and broad search query \textit{wikipedia.org has:links -is:retweet} with the two search operators \textit{has:links} and \textit{-is:retweet} to collect all original Tweets (no retweets) that included links to the domain \textit{wikipedia.org}. For running the data collection process, we used server instances with installations of R and RStudio on the interactive digital research environment UCloud\footnote{https://cloud.sdu.dk/app/dashboard} hosted by Syddansk Universitet's eScience Center. The data collection was performed over a period of five month in Summer 2021, and resulted in 44,945,098 Tweets posted between 2006 and January 2021. This collection of Tweets is what we call \texttt{TWikiL raw}. Based on this data, which contains all types of links to Wikipedia, we created a cleaned or curated version of this dataset, \texttt{TWikiL curated}, with Tweets that only contain links to Wikipedia article pages for which a Wikidata item could be retrieved. The necessary processing steps for the curated version are outlined below.

\textbf{Processing steps:} When processing the links and augmenting the retrieved Tweets we followed these steps:
\begin{itemize}
    \item We expanded shortened URLs.
    \item URLs that are redirects were resolved to the original article/redirect target by using the MediaWiki API\footnote{https://www.mediawiki.org/wiki/API:Redirects}.
    \item We performed URL encoding for all links to avoid character encoding issues.
    \item We used Wikidata's SPARQL query service\footnote{https://query.wikidata.org/} to retrieve Wikidata item identifiers associated with a Wikipedia article. For this step, additional cleaning of a Wikipedia URL was often necessary as Wikidata's query service does not recognise URLs from the mobile page\footnote{https://en.m.wikipedia.org/wiki/Twitter}, where the HTTP protocol is not secure (i.e. http only), or where users have been linking directly to a certain subsection of a Wikipedia article\footnote{https://en.wikipedia.org/wiki/Twitter\#Technology}.
\end{itemize}
We now continue with describing the dataset format.

\subsection{Dataset Format}
The SQLite database \texttt{TWikiL curated} contains the table \texttt{tweets\_urls} with the following twelve columns:

\begin{itemize}
    \item \textit{created\_at:} The timestamp for when a Tweet got created.
    
    \item \textit{tweet\_id:} The unique identifier of a Tweet.
    
    \item \textit{author\_id:} The unique identifier of the user that posted the Tweet.
    
    \item \textit{conversation\_id:} The unique identifier for a conversation. \textit{tweet\_id} and \textit{conversation\_id} are identical in case the Tweet did not receive a reply. In Twitters API v2 this identifier can be used to reconstruct conversations, i.e. retrieve direct replies and replies of replies for a certain Tweet ID.
    
    \item \textit{in\_reply\_to\_user\_id:} The user's ID for which the Tweet was a reply to. 
    
    \item \textit{tweet\_lang:} Language of the Tweet as detected by Twitter.
    \item \textit{reply\_binary:} This binary variable indicates whether the Tweet received a reply (1) or not (0).
    \item \textit{attention\_index\_scaled:} The values for the four engagement metrics reply, retweet, quote retweet and like, are summed up and scaled from 0 (no engagement) to 100 (Tweet with most engagement in the dataset). 
    \item \textit{wiki\_language\_edition:} Language code for the Wikipedia language edition the URL linked to. 
    \item \textit{wikidata\_id:} Wikidata identifier for the item/article that was referenced in the Tweet.
    \item \textit{wiki\_category:} The final two columns were created via a join with the dataset on Wikipedia knowledge propagation released by Valentim et al. \cite{Valentim2021}, who associated each Wikidata item to the ORES list of meta-topics or categories an item belongs to \cite{Wikipedia2021ORES}. A Wikidata item can be associated with multiple topics which are separated by a semicolon (;). 
    \item \textit{wiki\_score:}. Each topic listed in the previous column is associated to a value ranging from 0 to 1 indicating the likelihood of belonging to the topic. Only topics with a score of 0.5 or higher are mentioned in the previous column. 
\end{itemize}



\begin{table*}[h!]
\center
\begin{tabular}{@{}lllll@{}}
\toprule
                                                &  &  &                                 &  \\
\multicolumn{1}{r}{} &
   &
  \multicolumn{1}{c}{} &
  \multicolumn{1}{c}{\begin{tabular}[c]{@{}c@{}} \textbf{TWikiL curated} \\ \textbf{(N=35,252,782)}\end{tabular}} &
  \multicolumn{1}{r}{} \\ \midrule
\textbf{Tweet ID}                               &  &  &                                 &  \\
\multicolumn{1}{r}{N unique}                    &  &  & 34,543,612                      &  \\
\multicolumn{1}{r}{Min, Max, Mean (SD)}         &  &  & 1, 21, 1.02 (±0.22)             &  \\
\multicolumn{1}{r}{N link count \textgreater 1} &  &  & 474,577                         &  \\ \midrule
\textbf{Author ID}                              &  &  &                                 &  \\
\multicolumn{1}{r}{N unique}                    &  &  & 5,467,385                       &  \\
\multicolumn{1}{r}{Min, Max, Mean (SD)}         &  &  & 1, 586835, 6.4 (±449.22)        &  \\
\multicolumn{1}{r}{N tweets \textgreater 1}     &  &  & 2,318,883                       &  \\ \midrule
\textbf{Conversation ID}                        &  &  &                                 &  \\
\multicolumn{1}{r}{Part of conversation}        &  &  & 12,452,819 (35.3\%)             &  \\
\multicolumn{1}{r}{Not part of conversation}    &  &  & 22,799,963 (64.7\%)             &  \\ \midrule
\textbf{Reply to User ID}                       &  &  &                                 &  \\
\multicolumn{1}{r}{As reply}                    &  &  & 13,554,343 (38.4\%)             &  \\
\multicolumn{1}{r}{Not as reply (NA)}           &  &  & 21,698,439 (61.6\%)             &  \\ \midrule
\textbf{Tweet Language}                         &  &  &                                 &  \\
\multicolumn{1}{r}{N unique}                    &  &  & 66                              &  \\
\multicolumn{1}{r}{Min, Max, Mean (SD)}         &  &  & 60, 13398627, 534133 (±2033338) &  \\
\multicolumn{1}{r}{Top 3} &
   &
   &
  \begin{tabular}[c]{@{}l@{}}en: 13398627 (38.0\%)\\ ja: 9101602 (25.8\%)\\ und: 4356486 (12.4\%)\end{tabular} &
   \\ \midrule
\textbf{Reply}                                  &  &  &                                 &  \\
\multicolumn{1}{r}{Received reply}              &  &  & 6,722,393 (19.1\%)              &  \\
\multicolumn{1}{r}{Received no reply}           &  &  & 28,530,389 (80.9\%)             &  \\ \midrule
\textbf{Attention Index Scaled}                 &  &  &                                 &  \\
\multicolumn{1}{r}{Min, Max, Mean (SD)}         &  &  & 0, 100, 0.002 (±0.06)           &  \\
\multicolumn{1}{r}{N ais \textgreater 0}        &  &  & 11,706,639                      &  \\ \midrule
\textbf{Wikipedia Language Edition}             &  &  &                                 &  \\
\multicolumn{1}{r}{N unique}                    &  &  & 310                             &  \\
\multicolumn{1}{r}{Min, Max, Mean (SD)}         &  &  & 1, 19037434, 113719 (±1192793)  &  \\
\multicolumn{1}{r}{Top 3} &
   &
   &
  \begin{tabular}[c]{@{}l@{}}en: 19037434 (54.0\%)\\ ja: 8627176 (24.5\%)\\ es: 1891589 (5.4\%)\end{tabular} &
   \\ \midrule
\textbf{Wikidata ID}                            &  &  &                                 &  \\
\multicolumn{1}{r}{N unique}                    &  &  & 4,047,344                       &  \\
\multicolumn{1}{r}{Min, Max, Mean (SD)}         &  &  & 1, 50982, 8.71 (±82.68)         &  \\
\multicolumn{1}{r}{Top 3} &
   &
   &
  \begin{tabular}[c]{@{}l@{}}Q13580495: 50982 (0.15\%)\\ Q4351853: 34183 (0.1\%)\\ Q639444: 28433 (0.08\%)\end{tabular} &
   \\ \midrule
\textbf{Wikipedia Topic Category}               &  &  &                                 &  \\
\multicolumn{1}{r}{N unique}                    &  &  & 64                          &  \\
\multicolumn{1}{r}{N NA}                        &  &  & 6,379,849 (18.1\%)              &  \\
\multicolumn{1}{r}{Top 3} &
   &
   &
  \begin{tabular}[c]{@{}l@{}}Culture.Biography.Biography 7,500,625 (21.3\%)\\ STEM.STEM 6,796,897 (19.3\%)\\ Geography.Regions.Europe.Europe 4,594,853 (13.0\%)\end{tabular} &
   \\ \midrule
\textbf{Wikipedia Topic Category Score}         &  &  &                                 &  \\
\multicolumn{1}{r}{Min, Max, Mean (SD)}         &  &  & 0.5,1,  0.88 (±0.15)            &  \\
\multicolumn{1}{r}{N NA}                        &  &  & 6,379,849 (18.1\%)              &  \\ \bottomrule
\end{tabular}%
\caption{\label{tab:twikil-descriptive} Descriptive statistics of all variables in the dataset.}
\end{table*}


\subsection{Limitations}

During the dataset development process we were facing challenges that effect the degree to which this dataset can be characterized as the \textit{complete population} of all Tweets with Wikipedia links. 

\textbf{Processing challenges:} First, URLs in Tweets would not be extracted and expanded correctly if they are not complete (e.g. missing protocol), are broken (e.g. a missing character) or are glued to the rest of the Tweet text (e.g. due to a missing whitespace). Tweets with faulty URLs are still part of \texttt{TWikiL}, however, they were not caught up in the processing steps mentioned earlier, where we mostly relied on the URL entry in the \textit{url\_expanded} field of the Tweet entity object which would be empty in the mentioned cases. 

Secondly, the field \textit{url\_expanded} refers to an expansion of URLs shortened by Twitters custom URL shortening service, which has been active since June 2011 \cite{TwitterLinkShort2011}, and does not refer to cases in which users used an URL shortening service before including an URL in a Tweet. For example, a user might shorten the URL to the English Wikipedia article about Denmark via \textit{tinyurl.com} and integrate it in a Tweet. Twitter would additionally shorten this tinyURL via it's own \textit{t.co} shortening service, which results in a URL of the t.co format. The Tweet entity object field \textit{url\_expanded} contains the expanded version of the Twitter URL, i.e. the tinyURL version that still needs to be expanded fully, to reveal the actual Wikipedia URL. While Tweets with shortened URLs are still matching our query criteria and are retrieved in our search, we were left with a few shortened URLs that needed to be expanded. This was a time-consuming effort and also revealed some URL shortening services (e.g., p.tl or rusflat.com) that are not longer active today. However, the number of URLs shortened with defunct services is marginal and we have to note that for the majority of shortened URLs we were successful in expanding them and integrating them in \texttt{TWikiL curated}.

Thirdly, the process of retrieving Wikidata identifiers via sending Wikipedia URLs to Wikidata's SPARQL query service can be error prone. As mentioned before, multiple cleaning steps needed to be performed. We tried to resolve these issues, however, false negatives, i.e. article links for which the concept in the article has a Wikidata identifier, but wasn't retrieved by us due to errors in the URL sent to Wikidata's query service cannot be entirely excluded. Finally, Wikidata's SPARQL query service does, of course, not return an ID in cases where the URL/article did not exist at the time of posting (i.e. call to action) or has been deleted since then.

\textbf{Wikipedia article use beyond direct links:} We have to emphasise that our dataset only represents the metaphorical \textit{tip of the iceberg} when it comes to the entanglement and mutual relationship between Twitter and Wikipedia. First of all, as already mentioned above, \texttt{TWikiL curated} focuses on Wikipedia article links and especially the way Wikipedia content gets re-used. However, we are well aware that Wikipedia content can get reused in many other ways that go far beyond posting direct article links. This includes: (1) links to multimedia content, i.e. images, videos and audio files (2) direct text re-use with or without correct attribution or (3) screenshots of Wikipedia pages that get attached to a Tweet, which is very likely to occur in a mobile context. 

Nevertheless, \texttt{TWikiL} comes close to the complete population of Tweets with Wikipedia links and is one of the first datasets of it's kind that also takes a historical development into perspective. 

\begin{figure*}[ht]
    \centering
    \includegraphics[width=1\textwidth]{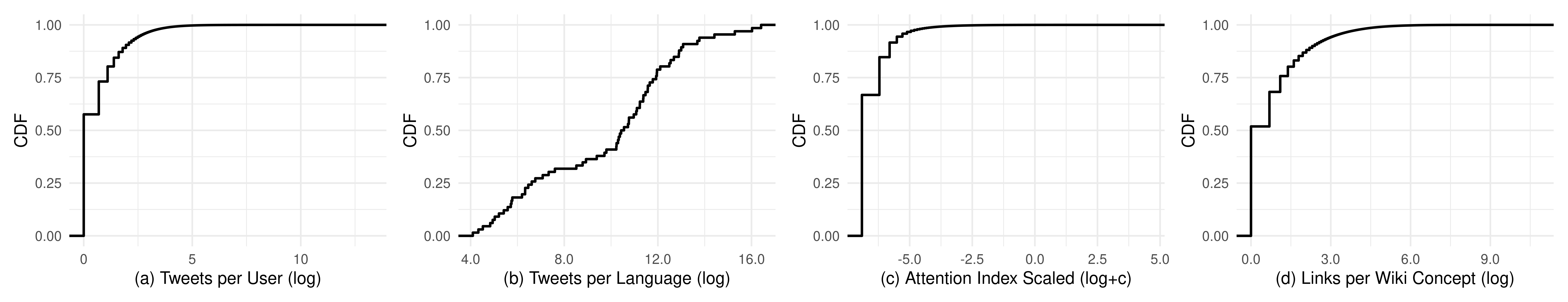}
    \caption{\label{fig:cdf-plots} Cumulative distribution functions (CDF) for (a) Tweets per User (log), (b) Tweets per Language (log) (c) Activity Index Scaled (log+c) (d) Links per Wikidata Item. In the case of Activity Index Scaled we added a constant $c=0.001$ to avoid infinite values at log transformation for Tweets that didn't receive any attention (0).}
\end{figure*}

\subsection{Ethical and FAIR Considerations}
\textbf{Personal data:} Neither \texttt{TWikiL raw} nor \texttt{TWikiL curated} contain personally identifiable information. While we further process the links included in Tweets and associate them with Wikidata identifiers, we do not derive or infer potentially sensitive characteristics about Twitter users or perform user-based off-Twitter matching, i.e. we do not associate "Twitter content, including a Twitter username or user ID, with a person, household, device, browser, or other off-Twitter identifier" \cite{Twitter2022}.

\textbf{FAIR principles:} The FAIR guiding principles for scientific data management and stewardship dictate that digital assests including datasets need to be findable, accessible, interoperable and reusable \cite{FAIRprinciples2016}. By making our dataset available via Zenodo \cite{Zenodo2013} we comply with most FAIR principles as they are practically inherent in the service \cite{Zenodo2022}. For example, via Zenodo \texttt{TWikiL} is assigned a digital object identifier (DOI), a globally unique and persistent identifier, and is findable and accessible via the services' search engine. The dataset is published in CSV (\texttt{TWikiL raw}) and SQLite (\texttt{TWikiL curated}) format respectively which are both public-domain and recommended storage formats by the Library of Congress \cite{SQLite2018}. The fact that every entry, i.e. link in our dataset, is associated with it's Wikidata identifier greatly enhances the dataset's interoperability and reusability. Researchers can easily join \texttt{TWikiL} with other Wikipedia and Wikidata related datasets that contain the Wikidata identifier which broadens the possible use cases and benefits the scientific community. To further enhance the datasets reusability, we published example code on how to make use of the dataset on Github \footnote{https://github.com/meier-flo/TWikiL}. The files found at this address exemplify how to work with the SQLite database from R and can be used to reproduce all descriptive statistics and visualisations presented in this paper. In addition to that, we provide skeleton code that allows to retrieve Tweet objects based on Tweet IDs using the R package \textit{rtweet} \cite{rtweet-package}. Via \textit{rtweet}, collecting Tweet objects can be done with a regular Twitter user account. A developer account is not necessary.

\section{Exploratory Analysis}

\begin{figure*}[ht]
    \centering
    \includegraphics[width=1\textwidth]{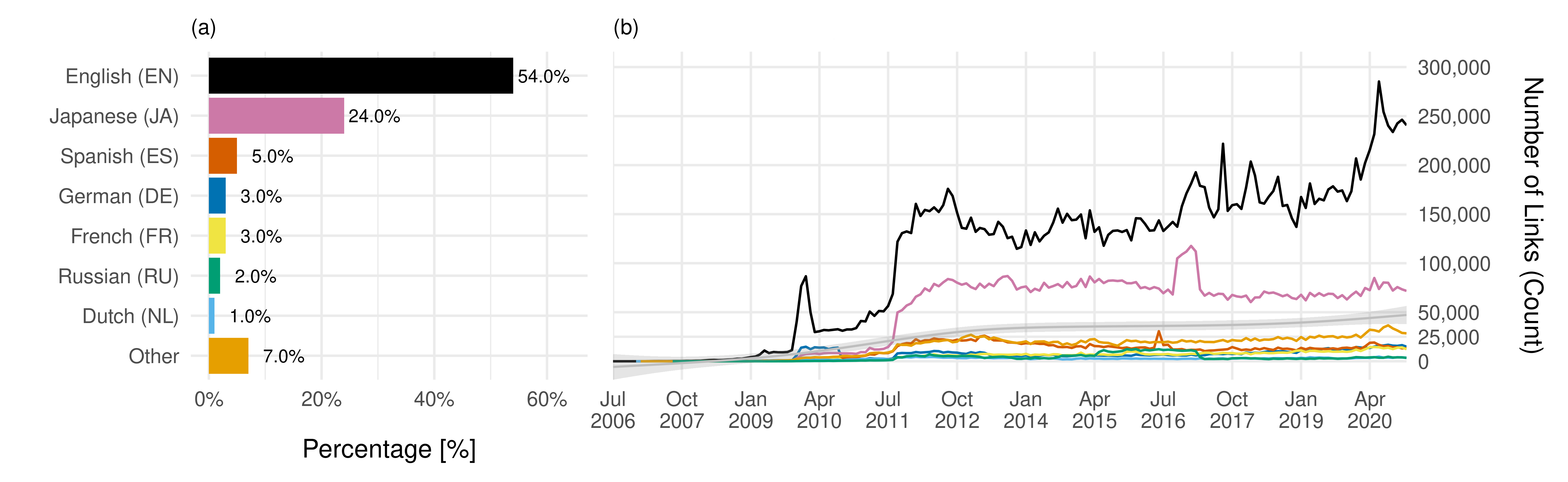}
    \caption{\label{fig:twikil_timeseries} (a) Share of the top seven most linked to Wikipedia language editions and (b) the monthly link count for these language editions over time. We picked the top seven language editions and a category \textit{Other}, as eight is the maximum number of colors for the color-safe OkabeIto color palette. The grey line in Figure (b) is a LOESS line indicating an overall increasing trend.}
\end{figure*}

We start our exploratory analysis of \texttt{TWikiL} by presenting and discussing descriptive statistics of some of the dataset's variables, all of which can be seen in Table \ref{tab:twikil-descriptive}.

\textbf{Tweet ID:} In total, around 5.5 M users posted 34.5 M Tweets linking to 35.3 M Wikipedia articles. In the dataset, we find 474,577 Tweets which link to two or more articles with one Tweet linking to 21 (Max) articles. A quick manual inspection of some of these Tweets shows that users motivation for doing this could be to highlight the differences or similarities of related concepts as in the case of Tweet ID \textit{143916372} between Q388-Linux and Q381-Ubuntu, or Tweet ID \textit{88843} between Q1190838-Pain au chocolat and Q207832-Croissant. 

\textbf{Author ID:} While we see 5.5 M different users adding Wikipedia links to their Tweets, the larger share (58\%) is doing this only once, which leads to an extremely long tailed distribution of Tweets per author (see \ref{fig:cdf-plots}(a)). On average an account posts 6.4 Tweets. 
The three accounts in the dataset that posted most Tweets with Wikipedia links are all bot accounts. It is very likely that bots play a major role in how Wikipedia links propagate through Twitter and future studies should investigate this further. 

\textbf{Conversation ID:} For around 12.5 M (35.3\%) database entries the Tweet ID and Conversation ID are not identical, which points to the fact that they are part of a discussion or conversation thread. 

\textbf{Reply to and Reply from:} A Tweet posted as reply has a Conversation ID that is different from its Tweet ID. However, the dataset contains instances where this is not the case. In total, there are around 1.1 M (3.1\%) Tweets that fall in this category. This results in differing percentages for the number of Tweets that are part of a conversation and the number of Tweets that are a reply to a user (13.6 M;38.4\%). Manual inspection showed that this is the case for account mentions that Twitters API classifies as replies but not as part of a conversation thread. At the same time, only about 20\% of Tweets with Wikipedia article links received a reply themselves. 

\textbf{Tweet Language:} The Tweet language is assigned by Twitter and part of the Tweet object returned by the API. In total 66 different languages were used. Most Tweets were written in English (38.0\%) or Japanese (25.8\%). 12.4\% of Tweets are classified as \textit{undefined} language. Figure \ref{fig:cdf-plots}(b) shows the distribution of languages indicating that the long tail is not as extreme as for other variables.  

\textbf{Attention Index:} The attention index is the scaled sum of all engagement values (reply, retweet, quote, like) so that the Tweet in the dataset with the highest engagement has a score of 100 (Max). The extremely low average (0.002) indicates that the largest part of Tweets with Wikipedia article links didn't receive much attention. In fact, 23.5 M or 66.8\% did not receive a retweet or similar. 

In what follows, we have a closer look at: (1) The volume of Wikipedia article links posted on Twitter with respect to differing language editions and time, (2) What meta topics these articles belong to and (3) on the most granular level, the most referenced Wikidata items. We have to note that especially (2) and (3) are language edition agnostic approaches as Wikipedia articles in different language editions have the same Wikipedia meta topic category and Wikidata ID.

\subsection{Volume of Wikipedia Article Links on Twitter}
Over the period of 15 years around 5.5 M users posted 35.3 M Wikipedia article links on Twitter. This is an average of 1974.1 links per month although link posting activity in Twitter's early years is only sparse. In 2006, for example, Twitters inauguration year, only 36 links were posted. Figure \ref{fig:twikil_timeseries}(a) and \ref{fig:twikil_timeseries}(b) give an overview on the share of links for the top seven language editions and their monthly development over time. Unsurprisingly, being by far the largest language edition with most readers and editors, more than half of all links posted on Twitter (54\%) are taken from the English language version. Links from the Japanese version account with 24\% for the second highest share followed by Spanish, German and French. Interestingly, links from Dutch Wikipedia account for the seventh highest share in the dataset (1\%) which seems remarkable given the fact that other language editions like the Italian, Arabic, Chinese, or Polish have much higher monthly page views. Table \ref{tab:twikil-descriptive} shows that in total users used 310 unique language editions although for many this was a single link. The average language edition is represented by 113,719 links whereby we can see an extreme variability in the dataset given the high standard deviation.
The grey slope in Figure \ref{fig:twikil_timeseries}(b) indicates an overall increasing trend for the volume of Wikipedia links on Twitter. In Twitter's early years not many Wikipedia links were posted. However, with
a growing user base the amount of shared Wikipedia links increased as well. December 2009 marked the first time that more then one hundred thousand article links (103,869) were posted per month. The most striking growth happened between June and September 2011, where both the number of English and Japanese Wikipedia links increased dramatically. This rise can also be noticed for other language editions, yet in a smaller dimension. A possible explanation for this development could be the release of Twitter's custom automatic link shortening feature, which was introduced in June 2011 and automatically shortens every link in a Tweet to a length of 23 characters \cite{TwitterLinkShort2011}. In other words, by increasing Twitter's usability and making it easier for users to post links likely resulted in a much higher usage of links in Tweets.
Moreover, in Figure \ref{fig:twikil_timeseries}(b) we can observe that while the number of Japanese article links stays fairly constant over time (except for the hump between October 2016 and February 2017), the number of English Wikipedia article links increased steadily, especially in the course of 2017 to now (Max=285,297 links in June 2020). However, this curve is also more volatile, which could be associated with public events leading to an increased interest in specific articles in general. This claim needs further investigations though.

\begin{figure*}[ht]
    \centering
    \includegraphics[width=1\textwidth]{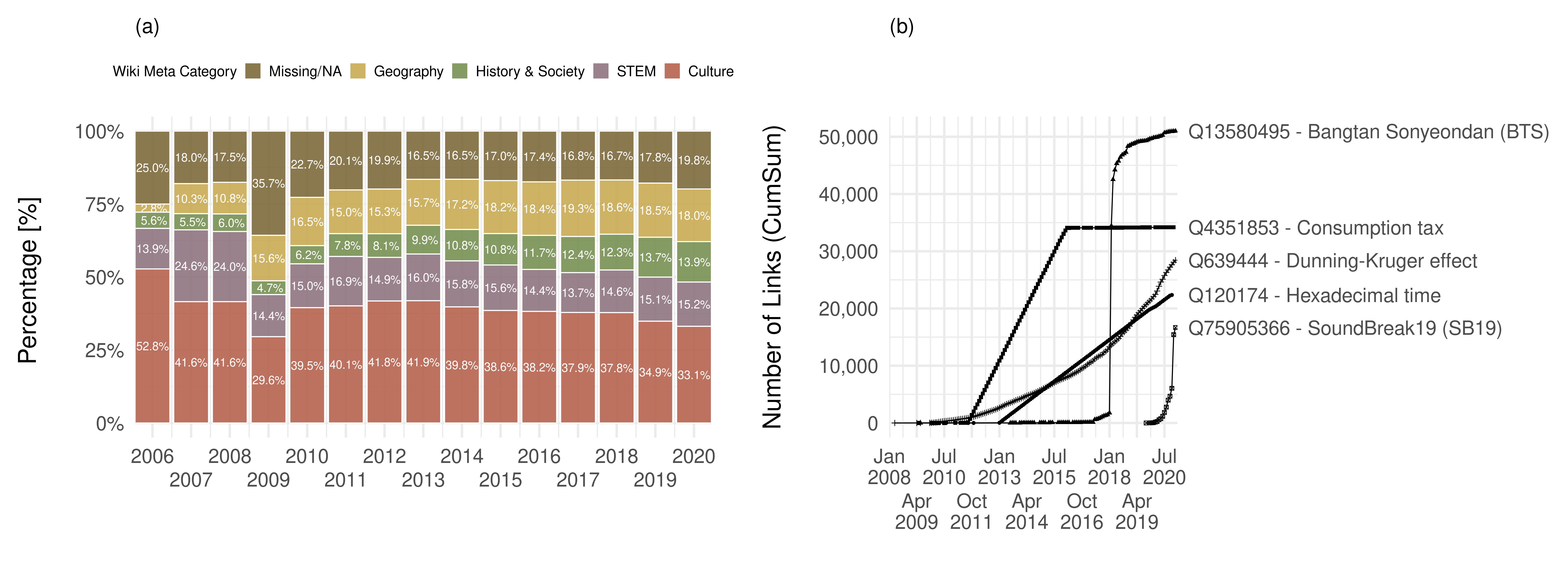}
    \caption{\label{fig:meta_cat_top5} (a) Share of links per Wikipedia meta topic including links with missing category entries per year. One has to note, that the number of absolute link count between 2006 (n=36) to 2020 (n=4,663,451) is increasing enormously. Figure (b) shows the cumulative sum of links for the top five most linked to concepts. Clear differences in propagation patterns can be observed. The linear increases for \textit{Q4351853-Consumption tax} and \textit{Q120174-Hexadecimal time} are hinting towards Tweets by bot accounts.}
\end{figure*}

\subsection{Wikipedia Meta Topics} 
In a next step, we were wondering how Wikipedia articles shared on Twitter are distributed over topics and whether one can observe any temporal variability in this distribution. To this end, we looked at which of the four meta topics or categories a posted link belongs to. For the sake of completeness, we add the category \textit{Missing/NA} which shows how many links are missing a category association. Figure \ref{fig:meta_cat_top5}(a) visualises the share per meta category posted per year. No extreme temporal dynamics are apparent: The ranking of article meta categories from most frequent to least frequent is Culture, Geography, STEM and History \& Society and this ranking does not change radically through the years. The popularity of Culture might be traced back to biography links which account for 21.3\% of all linked items (see Table \ref{tab:twikil-descriptive}). Although every fifth link to a Wikipedia article is a biography, the popularity of Culture articles is decreasing slightly, from 41\% of all articles in 2012 and 2013 to 33\% in 2022. Geography and History \& Society are slightly increasing, while STEM stays constant over the years. However, a more detailed analysis should be made that also looks at the relative number of articles per meta category available at these different point in times.

\subsection{Popular Concepts}
Finally, we zoom in even more by studying which Wikidata items are most popular and how their popularity developed over time. Figure \ref{fig:cdf-plots}(d) shows that more than half of all concepts were only posted once and that the distribution is again highly skewed. Among the top five most popular concepts we do not find historical figures or events as one could expect, but two boy bands, the South Korean boy band \textit{Bangtan Boys (BTS)} and the Filipino boy band \textit{SoundBreak 19 (SB19)}. While being among the most linked concepts they still account only for a very small percentage (see Table \ref{tab:twikil-descriptive}). To investigate how those concepts got popular over time, we created Figure \ref{fig:meta_cat_top5}(b) which shows the cumulative sum of item count for the top five concepts. Interestingly, both BTS and SB19 show a bursty development. For example, BTSs' popularity can be traced back to a surge in references to the concept in February 2018. A manual inspection of the Twitter users who posted most links to BTS articles using the OSoMe project Botometer\footnote{https://botometer.osome.iu.edu/} did not indicate that those accounts are social bots \cite{Davis2016}. However, the temporal development of two other concepts in the top five --- \textit{Q4351853-Consumption tax} and \textit{Q120174-Hexadecimal time} --- which follow a linear line, hints to the fact that an automated, steady link propagation is happening. For example, in the case of Hexadecimal time there are only 14 accounts tweeting about this item with one account being responsible for 99\% of all Tweets. Investigating concept propagation could thus help with identifying social bots too. While our analysis here is language agnostic, adding the Wikipedia language edition variable in the analysis would be an interesting next step. We believe that a closer look at what Wikidata items get posted, how they are used in Twitter conversations and how this develops over time, can say a lot about how our interests and web behaviour changes and how we develop as society. We will outline this use case further below.

\section{Potential Use Cases}
Due to it's unique nature, we believe that \texttt{TWikiL} offers a wide range of scholarly use cases. In the sections that follow, we present and discuss research questions and studies with respect to three areas: (1) Wikipedia's and Twitter's co-relationship (2) the use of Wikipedia links in Twitter conversations with a focus on fact-checking and fighting misinformation and (3) digital humanities and the history of the web.

\subsection{Wikipedia and Twitter - Twitter and Wikipedia}
\texttt{TWikiL} can be used to study the two platform's co-existence and codependency. In this context, multiple research questions and studies are imaginable. 
First of all, it would be interesting to study whether the frequency with which certain articles/concepts get posted on Twitter correlates with interest in Wikipedia articles and their pageview statistics, retrievable via the Wikimedia Pageviews API\footnote{https://wikitech.wikimedia.org/wiki/Analytics/AQS/Pageviews}. While interest in Wikipedia articles follows different temporal patterns depending on the topic, pageview statistics are also sensitive to current events. It is unclear whether seasonality and current events also lead to an increased propagation of articles with high pageview statistics to other platforms, or whether article propagation on Twitter follows different temporal patterns. Previous studies showed a direct link between sharing Wikipedia links on other platforms and increased Wikipedia article pageviews \cite{Moyer2015} and future studies can investigate this relation for Twitter too. To this end, the general motivation for posting Wikipedia articles should be studied and classified to get more insights in this information sharing behavior. In this scope, one could also study the \textit{community of practice} of Wikipedians on Twitter and how this community developed in terms of size and activities, as a better understanding of this community could also lead to better support in their practice as Wikipedia editors. 
Secondly, there is a need to study if the biases that Wikipedia articles suffer from with respect to the underrepresentation of certain geographical areas \cite{Graham2014}, gender \cite{Wagner2015}, or (non-western) culture \cite{Callahan2011,Ribe2018} also leads to an imbalance and bias when these articles get shared. Related inquiries with search engines showed that biased or one-sided Wikipedia links on SERPs greatly influences the searchers experience \cite{Vincent2019}. In this context the question is: can we identify whether these biases and imbalances also propagate to other platforms through sharing Wikipedia links? Studying cultural biases and diversity could be further supported by linking \texttt{TWikiL} to other existing datasets, for example, the Cultural Diversity Dataset \cite{Ribe2019}.
Thirdly, it is worth investigating if the paradox of re-use also applies in the context of Twitter, or, put another way, whether Tweets with Wikipedia links can work as motivator to create and/or edit Wikipedia content and thus be a driver for open collaboration. To this end, researchers could, for example, look into whether call-to action Tweets, in which users invite others to contribute to Wikipedia, lead to an increase in Wikipedia edits for these articles. A related question is whether the opposite can be observed and whether the distribution of Wikipedia links on Twitter also attracts users with malicious intentions which can lead to an increase in vandalism or even edit wars. Finally, the development of a complementary dataset, that collects all instances in which Tweets are used as references in Wikipedia articles would allow to further study Twitter and Wikipedia's co-dependency and give an even more holistic picture of their relationship. 

\subsection{Wikipedia Links in Twitter Conversations}
A second key thread of inquiry, from our point of view, is the way Wikipedia links get used in Twitter conversations. While conversation threads are also somewhat of importance in the previously mentioned use cases, here we specifically aim at the question of how Wikipedia links get used as reference, to support an argument, for fact checking and fighting misinformation and fake news in Twitter discussions. While designing and developing Wikipedia-based systems for fact checking is an active area of research in NLP \cite{Mykola2021,Bekoulis2021} and even large tech giants rely on human workforce and Wikipedia to check content on their platforms \cite{Flynn2017,Perez2020}, less is known about if, how and in which situations Wikipedia links are used by users for this purpose. Insights in this behaviour could also help with designing Twitter bots that mimic human behavior in sending Wikipedia links for fact-checking, possibly containing the spread of fake news.

\subsection{Digital Humanities and History of the Web}
Language is a proxy for culture and vice versa. In this sense, the dataset can be used to perform cross-cultural studies and analyse global digital cultures, by, for example, investigating Twitter users multilinguality and the use of different language editions when posting article links. Furthermore, the dataset can support studies about how certain user groups talk differently about cultural activities or cultural objects that the Tweets link to. By connecting \texttt{TWikiL} with other datasets, e.g. Valentim et al.'s dataset on knowledge propagation, comparison studies on how culture diffuses in social media or propagate on Wikipedia could be performed. \cite{Valentim2021}.

Finally, \texttt{TWikiL} can be considered a contribution to the field of \textit{web historiography}, which studies the history of the web and its development using mostly web archives as source \cite{Bruegger2019}. In this sense, \texttt{TWikiL} specifically focuses on the history of social media and the co-evolution and interwined genealogy of two of the most used social media platforms on the web \cite{Bruegger2018}. While \texttt{TWikiL} does not have archive character in the most general sense, as we cannot guarantee that Twitter will always give access to data via an API as it does now. However, as long as the API stays unchanged, the dataset solves one of the main challenges mentioned by web historians which is the desideratum for publicly accessible datasets that allows the study of social media in its closely integrated form. Due to its unique nature, \texttt{TWikiL} gives access to this historical and integrated perspective. 

\section{Conclusion}
As we see more and more evidence for web platforms being tightly connected to each other, there is an increasing need for not only studying them in isolation but paint a holistic picture of their dependency and the value they contribute to each other. For this purpose we developed \texttt{TWikiL}, a dataset containing all Wikipedia article links posted on Twitter between 2006 and January 2021. The dataset enables a wide range of scholarly use cases for studying Twitter and Wikipedia's interrelation, but also for the digital humanities and the history of the web. Initial exploratory analysis shows how increased user experience can contribute to higher information diffusion and the importance of culture-related articles --- especially biographies --- for a social media platform. 
In addition to the data, we also release code that gives insights in how to work with the dataset, to get fellow researchers started with using \texttt{TWikiL} in their own research.

\bibliographystyle{aaai}
{\small
\bibliography{wiki-ref.bib}}
\end{document}